\documentclass[a4paper,fleqn,usenatbib]{mnras}

\usepackage[T1]{fontenc}
\usepackage{ae,aecompl}
\usepackage{soul}

\usepackage{graphicx}	
\usepackage{amsmath}	
\usepackage{amssymb}	

\usepackage{mathtools}

\usepackage[usenames, dvipsnames]{color}

\title[Signatures of migrating planets]{Revealing signatures of planets migrating in protoplanetary discs with ALMA multi-wavelength observations}

\author[P. Nazari et al.]{Pooneh Nazari,$^{1}$\thanks{E-mail: pn297@cam.ac.uk}
Richard A. Booth,$^{1}$\thanks{E-mail: rab200@ast.cam.ac.uk}
Cathie J. Clarke,$^{1}$
Giovanni P. Rosotti,$^{1,2}$\newauthor
Marco Tazzari,$^{1}$
Attila Juhasz,$^{1}$
and Farzana Meru$^{3,4}$
\\
$^{1}$Institute of Astronomy, University of Cambridge, Madingley Road, Cambridge, CB3 0HA, UK\\
$^{2}$Leiden Observatory, University of Leiden, P.O. Box 9500, NL-2300 RA, Leiden, the Netherlands\\
$^{3}$Department of Physics, University of Warwick, Gibbet Hill Road, Coventry CV4 7AL, UK \\
$^{4}$Centre for Exoplanets and Habitability, University of Warwick, Gibbet Hill Road, Coventry CV4 7AL, UK
}

\date{Accepted 2019 March 15. Received 2019 March 7; in original form 2019 January 23}

\pubyear{2019}

\begin{document}
\label{firstpage}
\pagerange{\pageref{firstpage}--\pageref{lastpage}}
\maketitle

\begin{abstract}
Recent observations show that rings and gaps are ubiquitous in protoplanetary discs. These features are often interpreted as being due to the presence of planets; however, the effect of planetary migration on the observed morphology has not been investigated hitherto. In this work we investigate whether multiwavelength mm/submm observations can detect signatures of planet migration, using 2D dusty hydrodynamic simulations to model the structures generated by migrating planets and synthesising  ALMA continuum observations at $\rm 850\,\micron$ and $\rm 3\,mm$. We identify three possible morphologies for a migrating planet: a slowly migrating planet is associated with a single ring outside the planet's orbit, a rapidly migrating planet is associated with a single ring inside the planet's orbit while a planet migrating at intermediate speed generates one ring on each side of the planet's orbit. We argue that multiwavelength data can distinguish multiple rings produced by a migrating planet from other scenarios for creating multiple rings, such as multiple planets or discs with low viscosity. The signature of migration is that the outer ring has a lower spectral index, due to larger dust grains being trapped there. Of the recent ALMA observations revealing protoplanetary discs with multiple rings and gaps, we suggest that Elias 24 is the best candidate for a planet migrating in the intermediate speed regime. 

\end{abstract}




\section{Introduction}

High resolution images of protoplanetary discs have unveiled a wealth of structures, such as gaps, rings and cavities in the (sub)-millimeter continuum emission with the Atacama Large Millimeter Array (ALMA) and scattered light via the Spectro-Polarimetric High-contrast  Exoplanet  REsearch  (SPHERE) and the Gemini Planet Imager (GPI) instruments (e.g. \citealt{ALMAPart2015}; \citealt{Rapson2015}; \citealt{Clarke2018}; \citealt{SPHERE2019}, \citealt{Beuzit2008}, \citealt{Macintosh2014}), with spirals and point sources also present in scattered light images (e.g. \citealt{Benisty2015}; \citealt{Wagner2015}). The origin of these structures is much debated, but  points either to some disc process (e.g. associated with magnetohydrodynamical effects, \citealt{Flock2015}, or opacity effects at  ice lines, \citealt{Zhang2016}, though see \citealt{Long2018}), or the presence of planets. The association of \emph{some} of these structures with planets is almost certainly true, given that direct imaging surveys have found planets in similar ranges of semi-major axis \citep{Marois2008,Biller2013,Vigan2017,Keppler2018}. 

A number of studies have suggested that these structures contain more information than just the presence of planets. For example, gap depths and widths both in the dust and the gas surface density profile can help to constrain the planet mass \citep{Crida2006,Kanagawa2015,Rosotti2016,Zhang2018}. \cite{Teague2018} argue that the deviations in the rotation curve due to planets can be derived from CO line observations, placing similar constraints on the mass. Furthermore, the gap structure can place constraints on the viscosity in discs because planets can open multiple gaps when the viscosity is low \citep{Dong2018,Bae2018a}. \citet{Fedele2018} argued that this is likely the case in AS 209, with one planet opening 2 or 3 gaps. Recent high resolution data from the DSHARP survey support this idea \citep{Andrews2018,Guzman2018,Zhang2018}. Another idea is to use the width of rings in ALMA images to constrain the strength of turbulence \citep{Dullemond2018}.

Recently, \citet{Meru2018} argued that planet migration may have an effect on the observed morphology of gaps and rings opened by planets. This is based on the idea that the planet's radial velocity relative to the radial velocity of dust in the planet's vicinity determines where dust particles concentrate. If the planet is migrating slowly with respect to the dust, one obtains the morphology well known in the context of transition discs \citep{Rice2006}. In this case dust from large distance drifts towards the planet, being trapped outside its orbit, while dust inside the planet's orbit drifts towards the star, leaving behind a cavity. \citet{Meru2018} showed that a different morphology arises when the planet migrates faster than the dust drifts. In this case the planet sweeps up the dust ahead of it, creating a ring inside the planet's orbit and leaves the dust outside of the planet's orbit behind.

The idea that ALMA observations can constrain the migration speed of planets in discs could have a tremendous impact on our understanding of planet formation and evolution. This is because, under the assumption that the structures are due to planets, the inferred planet masses and semi-major axes (typically in the Neptune to Jupiter range at radii of 10s of au) sample a region of parameter space that is hard to access via traditional exoplanet detection methods (i.e. radial velocity and transit method are suited to finding planets within a few au, while direct imaging surveys, which target larger semi-major axes, are restricted by sensitivity limits to masses above a few times Jupiter's mass). To date,  while planetary migration is widely invoked as a theoretical concept and as a key ingredient in determining planetary system demographics, there has been no direct observational confirmation of the process.

In this paper we analyse the morphological imprint of planet migration. To this end, we consider multiwavelength continuum brightness profiles, taking into account the full range of dust grain sizes present. This enables us to explore how the picture put forward by \citet{Meru2018} is affected by the complication of having different dust grain sizes, which travel at different speeds. In \autoref{sec:method} we conduct 2D dusty hydrodynamic simulations and produce synthetic ALMA observations to explore whether the structures remain observable when taking into account the full distribution of grain sizes present in the disc. In \autoref{sec:param_space_explained} we present the physical basis for the different results, summarizing the key results in \citet{Meru2018}. In \autoref{sec:profiles} we examine the synthetic brightness profiles for different choices of planet migration speed. We discuss the circumstances under which these signatures can be distinguished from other mechanisms that produce gaps and rings. 

Finally, in \autoref{sec:discussion} and \autoref{sec:conclusion} we discuss our results in the context of observed systems and present our conclusions.

\section{Methods}
\label{sec:method}
\subsection{Hydrodynamic simulations}
\label{sec:num_method} 
We conduct multi-species gas and dust simulations of the gap structure formed by migrating planets, in the same way as \citet{Meru2018}. We use the code FARGO3D \citep{fargo3d2016}, running in a 2D cylindrical geometry, which we adapted to include dust dynamics \citep{Booth2015, Rosotti2016}. A logarithmic grid is used,  with radius extending from 0.1 to 3 times the planet's initial radius, comprising   1257 and 680 cells in the azimuthal and radial directions respectively. The gas surface density is set to
\begin{equation}
  \Sigma = \Sigma_{0} \left(\frac{R_{0}}{R}\right),
\end{equation}
where $R$ is the radius in the disc, $R_{0}$ is the initial location of the planet and $\Sigma_{0}$ is the gas surface density at $R_{0}$. The disc is modelled as locally isothermal with a constant flaring index, where the aspect ratio, $h$, is
\begin{equation}
  h = \frac{H}{R} = h_{\rm p}\left(\frac{R}{R_{0}}\right)^{0.25},   \label{HoverR}
\end{equation}
where $H$ is the pressure scale height, $h_{\rm p}$ is the aspect ratio at the planet's initial location. The parameters used are shown in \autoref{tab:mig}. We include viscosity via the $\alpha$ prescription of \cite{visc1973} and set $\alpha_{\nu} = 10^{-3}$. 

We use 13 dust sizes in the simulations with logarithmically spaced Stokes numbers between 0.1 and $\rm 10^{-5}$. The maximum Stokes number is chosen to be consistent with  the maximum size in grain-growth models \citep{Birnstiel2012}. The minimum Stokes number is chosen so that it is 100 times smaller than $\alpha_{\nu}$ because for dust sizes smaller than this the dust follows the gas (\citealt{Takeuchi2002,Jacquet2012}). 

The planets are initially located at 1 code unit in the disc and allowed to migrate immediately. We model planets with $M_{\rm p} = 30\, \rm M_{\oplus}$ and prescribe the migration  as in \cite{Meru2018},
\begin{equation}
  R_{\rm p} = R_{0}e^{-t/\tau_I},
\end{equation}  
where $R_{\rm p}$ is the planet's location, $t$ is the time after the planet is released, and $\tau_I$ is the Type I migration timescale, which is given by
\begin{equation}
  \tau_{I} = \frac{R_{\rm p}^2\Omega_{\rm p}M_{\rm p}}{2|\Gamma_{\rm p}|}, \label{eqn:typeImig}
\end{equation}
where $\Omega_p$ is the Keplerian frequency found at the location of the planet. The torque exerted by the disc on the planet, $\Gamma_{\rm p}$, is given by 
\begin{equation}
  \label{torque}
  \Gamma_{\rm p} = - 1.95\left(\frac{q}{h}\right)^2 \Sigma_{\rm p} R_{\rm p}^4 \Omega_{\rm p}^2.
\end{equation}
\citep{Baruteau2014}. Here $q$ is the planet-to-star mass ratio. \autoref{tab:mig} shows a subset of the simulations that we have run with the planet and disc parameters used. 

\begin{table*}
	\begin{centering}
	\caption{The characteristics of the models whose intensity profiles are shown in \autoref{graph:multi_int} and \autoref{graph:multi_int_1cm}. In all the simulations mass of the planet is 30\,$\rm M_{\oplus}$. The first column shows the migration time in orbits calculated using the orbital period of the planet at its initial location. The second column shows the location of the planet after migrating in au, the third column shows the initial location of the planet in au, the fourth column shows the gas mass
	derived from the migration time scale (\autoref{eqn:typeImig}), the fifth column shows the critical dust grain size, the sixth column shows the critical Stokes number and the last two columns show $\rm h$ and $\rm h_{au}$ as defined in \autoref{HoverR} and \autoref{eq:temp}.}
	
	\label{tab:mig}
	\begin{tabular}{lcccccccr} 
	  \hline
          \hline
		$\rm Model$ & $\tau_I$\,$\rm (orbits)$ & $\rm R_p$\,$\rm (au)$ & $\rm R_0$\,$\rm (au)$ & $\rm M_{disc}$\,$(\rm M_{\odot})$ & $a_{\rm crit}$\,$\rm (\mu m)$ & $\rm St_{\rm crit}$ & $\rm h$ & $\rm h_{\rm au}$\\
		\hline     
		Slow          & 2865 & 54.17 & 71.6 & $2.418 \times 10^{-3}$ &  49 & 0.012 & 0.045 & 0.0165 \\
		Intermediate  &  975 & 38.71 & 71.6 & $5.076 \times 10^{-3}$& 301 & 0.026 & 0.041 & 0.0165 \\
		Fast          &  716 & 40.97 & 71.6 & $7.315 \times 10^{-3}$& 591 & 0.037 & 0.042 & 0.0165\\
		\hline
                \hline
	\end{tabular}\\
        \end{centering}
        \begin{flushleft} 
        \end{flushleft}  
\end{table*}

\subsection{Radiative transfer}
\label{sec:calc_int}
To investigate whether the signatures of migration are observable at (sub)-millimetre wavelengths, we compute the dust continuum emission assuming it comes from the mid-plane using the same temperature profile as in the hydrodynamical simulations, i.e. we convert the disc aspect ratio (\autoref{HoverR}) into temperature assuming a mean molecular weight of 2.3 and a $\rm 1\,M_{\odot}$ star, giving
\begin{equation}
T = 100 \left(\frac{h_{\rm au}}{0.02}\right)^2 \left(\frac{R}{1\,{\rm au}}\right)^{-0.5} \,{\rm K},
\label{eq:temp}
\end{equation}
where $h_{\rm au}$ is the disc aspect ratio at 1~au. The intensity is given by
\begin{equation}
  \label{eq:intensity}
  I_{\nu} = B_{\nu}(T) (1 - e^{-\tau_{\nu}}),
\end{equation}
where $B_{\nu}(T)$ is the Planck function at temperature $T$ and $\tau_{\nu}=\displaystyle \sum_{a}\Sigma_{a} \kappa_{a}$ is the optical depth at frequency $\nu$. In the expression for optical depth, $\Sigma_{a}$ and $\kappa_{a}$ are the surface density and opacity of dust with size $a$ respectively. In \autoref{subsec:max_dust} we show the optical depth for four ranges of dust sizes using this expression. 

The optical depth is computed assuming an opacity of compact spherical grains, found from Mie theory and using the optical constants of astronomical silicates (\citealt{opacity2003}).  

The dust surface density for each grain size is calculated in the same manner as in \citet{Rosotti2016}. We take the dust surface densities from the hydrodynamic simulations and convert them to dimensional form by assuming an initial dust-to-gas ratio of 0.01 to normalise the densities. We assume 66 logarithmically spaced dust grain sizes between $0.5\,\rm \mu m$ and a maximum grain size of 1\,mm or 1\,cm, where the grains follow a size distribution of ${\frac{dN}{da}} \propto a^{-3.5}$. 

The Stokes number is given by 
\begin{equation}
  \label{stokes_a}
  \textup{St} = t_{\rm s} \Omega = \frac{\rho_{\rm d}\,a\,\Omega}{\rho_{\rm g}\,C_{\rm s}}\sqrt{\frac{\pi}{8}} = \frac{\pi}{2}\frac{\rho_{\rm d}\,a}{\Sigma_{\rm g}},
\end{equation}
where $t_{\rm s}$ is the stopping time, $\rho_{\rm d}$ is the bulk density of the dust which we have assumed to be $\rm 1\,g\,cm^{-3}$, $\rho_{\rm g}$ is the mid-plane gas density and $\Sigma_{\rm g} = \sqrt{2\pi}\,H\,\rho_{\rm g}$ is the gas surface density. For comparison, grains with Stokes number of 0.1 will have size 
\begin{equation}
a_{\rm max} = 0.9 \times  \left(\frac{M_{\rm disc}}{0.005\,{\rm M_{\odot}}}\right) \left(\frac{R_{\rm p}}{50\,{\rm au}}\right)^{-1}   \,{\rm mm},
\end{equation}
where a disc radius of 100 au has been assumed. 

To produce the 1D brightness profiles of the disc shown in \autoref{sec:profiles}, we azimuthally average the dust surface densities at each radius and use this average in \autoref{eq:intensity}. Note that for the planet mass considered here, no spirals are  visible in the simulated images.

\subsection{Simulated observations}

We also produce synthetic ALMA observations for a subset of our simulations to test which features are observable in realistic observations. We compute ALMA synthetic observations at $\rm 850\,\mu m$ (band 7) and $\rm 3\,mm$ (band 3) using the `simobserve' tool in Common Astronomy Software Applications package (CASA) (\citealt{casa2007}). At each wavelength we combine an extended configuration and a compact configuration of ALMA's 12\,m Array to sample all the spatial scales present in the simulations between $0.028\arcsec$ and $1.5\arcsec$. Configurations C43-8 and C43-5 are combined for $\rm 850\,\mu m$ (352.71\,GHz) and configurations C43-9 and C43-6 are combined for $\rm 3\,mm$ (99.93\,GHz). The beam size is $0.034\arcsec \times 0.031\arcsec$ for the $\rm 850\,\micron$ emission and $0.067\arcsec \times 0.056\arcsec$ for the $\rm 3\,mm$ emission. We use integration times of 2 hours and 5 hours for C43-8 and C43-9 respectively. We find the integration times for configurations C43-5 and C43-6 by multiplying the integration times for C43-8 and C43-9 by 0.22 and 0.21 multipliers, as advised by the ALMA Proposer Guide. We assume a continuum bandwidth of 7.5\,GHz at each wavelength. We assume the disc is face-on and is located at a distance of 140\,pc.  

The outputs of `simobserve' are cleaned using the CASA `tclean' task (\citealt{clean2017}). We cleaned the images down to the noise levels. The noise in Band 7 is $\rm 1.85 \times 10^{-2}\,mJy/beam$ and in Band 3 is $\rm 4.79 \times 10^{-3}\,mJy/beam$. We azimuthally average the ALMA synthetic image at each radius to produce the 1D brightness profiles of the synthetic observations shown in \autoref{sec:profiles}. The errors on the azimuthally averaged brightness profiles reported in \autoref{graph:multi_int} and \autoref{graph:multi_int_1cm} not only include the sensitivity limit of the observations (as computed by the ALMA simulator), but also account for the number of independent brightness measurements obtained at each radius, $\rm \sqrt{\frac{2\pi R}{b_{\rm maj}}}$, where $\rm b_{\rm maj}$ is the beam's semi-major axis.

To obtain unbiased radial profiles of the spectral indices, we use the same range of baselines ($\rm 17.65\,k\lambda$-$\rm 4630\,k\lambda$) and the same restoring beam ($\rm 0.067\arcsec \times 0.056\arcsec$). Then we azimuthally average the data at each radius and find the errors as described above. Next, we calculate the spectral indices using the following equation

\begin{equation}
    \alpha = \frac{\log{(I_7/I_3)}}{\log{(\nu_{7}/\nu_{3})}},
    \label{eq:spectral_index}
\end{equation}
where $\nu$ is the frequency and the subscripts indicate the ALMA bands. Finally, we propagate the errors in intensities to the error in spectral index starting from \autoref{eq:spectral_index}.    

\section{Model parameters}
\label{sec:param_space_explained}

As shown in \citet{Meru2018}, the morphology of the dust surface density depends on the relative radial velocity of the dust and planet. Large dust grains can migrate faster than the planet, meaning that dust grains from large distances become trapped in the pressure maximum outside it's orbit, while large dust grains inside the planet's orbit run away towards the star. Conversely small dust grains can migrate more slowly than the planet, leading to the opposite behaviour -- i.e. dust inside the planet's orbit is caught up and swept up into a ring inside the planet's orbit, while small dust grains outside the planet's orbit are left behind. This relies on the planet opening up a deep enough gap that the dust cannot pass through it, i.e. $M_p \gtrsim 20\,M_\oplus$ \citep{Rosotti2016}. Thus we focus on the case of a $30\,M_\oplus$ planet, which opens up enough of a gap to prevent dust flowing through its orbit without being so massive as to be in the Type II migration regime, where the planet is expected to migrate much slower than the dust.

ALMA offers the best chance of observing such an effect. This is due to the combination of several factors: 1) the disc is sufficiently optically thin  that the emission traces the mid-plane density 2) the spatial resolution is sufficient to probe planet-induced gaps and 3) the parameter space where the migration speed of a planet is comparable to the speed of the dust lies in a range that produces observable features at mm wavelengths. We therefore focus on grain sizes in the vicinity of 1~mm.

To determine the parameter space for suitably sized dust we can compare the velocity of the planet, $v_{\rm p}$, to $v_{\rm d}$, the velocity of the dust \citep{Meru2018}. These are given by 
\begin{equation}
v_{\rm p} = \frac{R_p}{\tau_I},
\end{equation}
and 
\begin{equation}
  \label{v_dust}
  v_{\rm d} = \frac{St}{\Sigma_g \Omega}\frac{\partial P}{\partial R},
\end{equation}
where the 2D expression for the radial drift velocity is used to be compatible with the numerical simulations (the 3D expression is only different by a factor of order unity, c.f. \citealt{Takeuchi2002}). Combining these with \autoref{eqn:typeImig} we find the Stokes number and grain size at which  $v_{\rm p} = v_{\rm d}$ for a given disc mass and radius:
\begin{align}
    \label{eqn:crit_st}
    {\rm St_{crit}} \sim&~0.015\left(\frac{M_{\rm p}}{30\,{\rm M_{\oplus}}}\right) \left(\frac{M_{\rm disc}}{0.005\,{\rm M_{\odot}}}\right) \left(\frac{R_{\rm disc}}{100\,{\rm au}}\right)^{-1} \notag \\
    & \times \left(\frac{h}{0.05}\right)^{-4}\left(\frac{M_{\star}}{1\,{\rm M_{\odot}}}\right)^{-2} \left(\frac{R_{\rm p}}{50\, {\rm au}}\right) ,
\end{align}

\noindent and
\begin{align}
  \label{eqn:crit_a}
  a_{\rm crit} \sim&~0.13 \left(\frac{M_{\rm p}}{30\,{\rm M_{\oplus}}}\right) \left(\frac{M_{\rm disc}}{0.005\,{\rm M_{\odot}}}\right)^2 \left(\frac{\rho_{\rm d}}{1\,{\rm g\,cm^{-3}}}\right)^{-1}
  \notag \\& \times \left(\frac{R_{\rm disc}}{100\,{\rm au}}\right)^{-2}  
  \left(\frac{M_{\star}}{1\,{\rm M_{\odot}}}\right)^{-2} \left(\frac{h}{0.05}\right)^{-4} {\rm mm}.
\end{align}

In the following sections we will compare a given grain size to the critical grain size as defined above. In particular we will compare $a_{crit}$ both with the minimum grain size that contributes significantly to the emission at wavelength $\lambda$, i.e. $2\upi a \sim  \lambda$ \citep{Draine1984}, and also the maximum grain size present (1~mm or 1~cm).

\section{Results}   
\label{sec:profiles}

We have explored the range of morphologies using simulations with different migration speeds and disc aspect ratios. We first examine three simulations that demonstrate the range of behaviours seen in subsections \ref{first_order_intensity} to \ref{sec:spec_index}, before presenting our exploration of the parameter space in subsection \ref{sec:parameters_of_disc}.

\subsection{Morphology}
\label{first_order_intensity} 

We begin by discussing the morphology of the continuum emission for the case of a slowly migrating planet, $v_{\rm p} \ll v_{\rm d}$, an intermediate case, $v_{\rm p} \approx v_{\rm d}$, and a fast case, $v_{\rm p} \gg v_{\rm d}$. As discussed above, the regime that the simulation occupies depends on both the grain size and the migration rate of the planet, so it is necessary to decide upon a grain size. In the discussion below, we label the three simulations based upon which regime they would be in when $a_{\rm max} = 1\,{\rm mm}$. Note that the morphology is dependent on  the simulation regime which itself depends on the choice of maximum grain size (see \autoref{subsec:max_dust}).

\begin{figure*}
  \centering
  \includegraphics[width=0.8\textwidth]{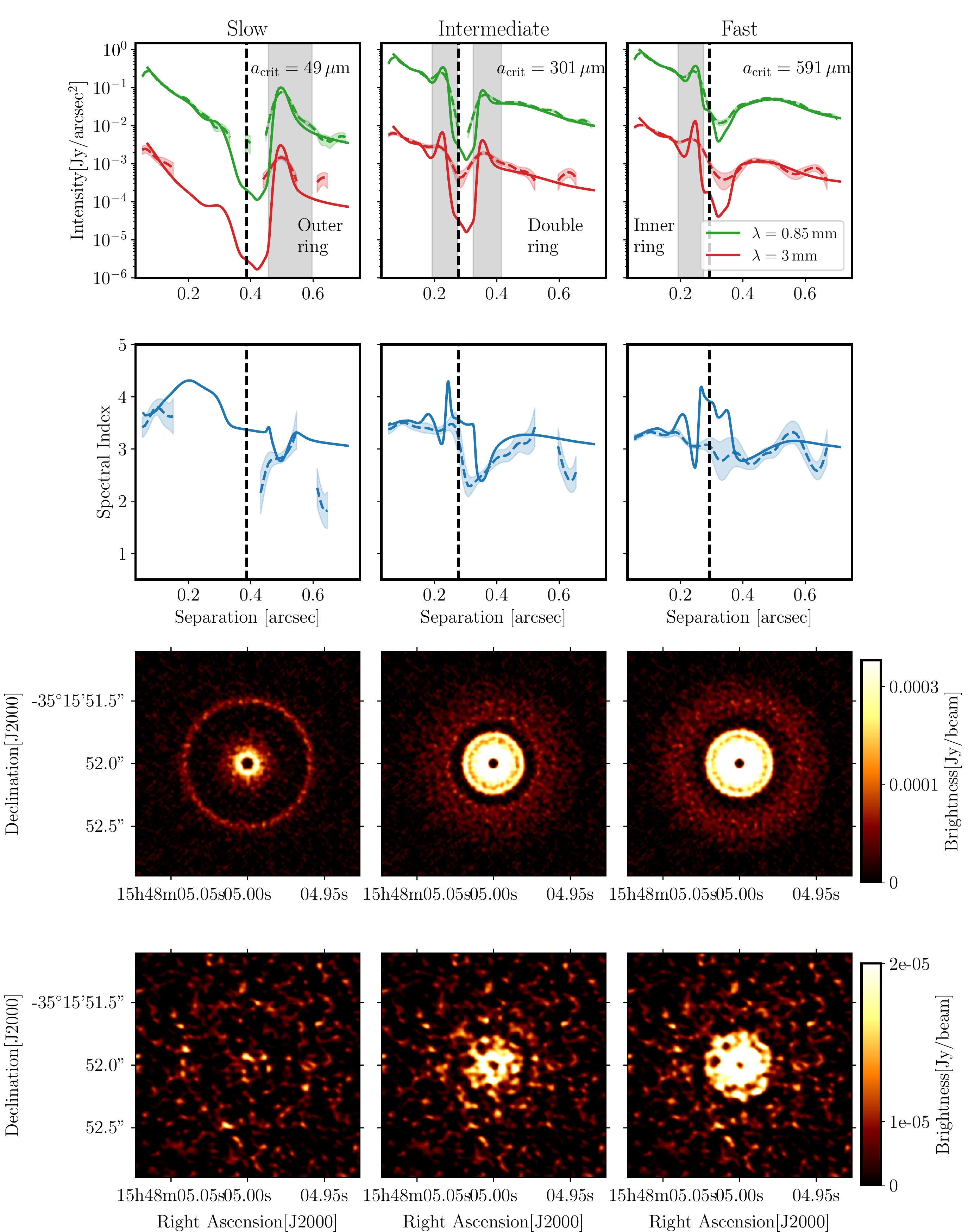}
  \caption{Simulated intensity profiles for the three simulations in \autoref{tab:mig}, assuming the maximum grain size is 1~mm. Top row: Red and Green solid lines show the intensity profiles from the FARGO3D simulations, the corresponding dashed lines show the profiles derived from the simulated ALMA images with the shaded red and green areas around them indicating the $1\sigma$ errors (regions with signal-to-noise $<2$ have been excluded from the profiles), the black dashed line shows the location of the planet and the shaded grey areas indicate the rings or dust enhancements. Second row: The blue solid and dashed lines show the spectral indices from the hydrodynamical simulations and the simulated ALMA images with the shaded blue areas around the dashed lines indicating the $1\sigma$ errors, the black dashed line shows the location of the planet. Third Row: Simulated ALMA images at 850~\micron. Bottom Row: Simulated ALMA images at 3~mm.}
  \label{graph:multi_int}
\end{figure*}

The simulated brightness profiles, along with the synthetic images, reconstructed brightness profiles, and spectral indices are shown in \autoref{graph:multi_int}.  The parameters of these three models shown are given in \autoref{tab:mig}.

We  first consider the two extreme cases, the slow planet case ($v_{\rm p} \ll v_{\rm d}$) and the fast planet case ($v_{\rm p} \gg v_{\rm d}$). The slow planet case, with $a_{\rm crit} = 40\micron$, is characterised by a single ring outside the planet's orbit and a low surface brightness inside the planet's orbit because most of the dust inside the planet's orbit has drained onto the star. 

The fast planet case instead shows a single ring inside the planet's orbit. This essentially confirms the behaviour identified by \citet{Meru2018}, where the position of the bright ring relative to the planet depends on the migration speed. This shows that both morphology types can be identified by ALMA (for a further discussion see \autoref{sec:spec_index}). We note that in the fast case we have $a_{\rm crit} \approx 600 \micron$, so there is some dust migrating faster than the planet. However, the contribution to the opacity in Band 7 from dust with $v_{\rm d} > v_{\rm p}$ is sufficiently small that the morphology is not considerably affected by the presence of these grains. In Band 3 the largest grains have a more important contribution to the opacity, which explains why the outer edge of the gap is more well defined in the longer wavelength observation. However, even in this case there is no clear bright ring outside of the planet, with the main bright ring appearing just inside the gap.

The intermediate case is characterised by two rings, one either side of the planet's orbit, which can be seen at $850\,\micron$ in \autoref{graph:multi_int}. In this case we have $a_{\rm crit} \approx 0.3\,{\rm mm}$, such that the amount of mass  trapped outside the planet's orbit is comparable to the amount swept up by the planet on the interior side. 

The slow case has a similar morphology to a transition or pre-transitional disc \citep[e.g.][]{Espaillat2007}, i.e the surface brightness inside the gap is much lower than outside it. For faster migrating planets the surface brightness inside the planet's orbit is instead comparable or higher than outside the planet's orbit.
This  suggests that there is a combination of characteristics (an exterior ring {\it and} a bright inner disc) which is an unambiguous signature of planetary migration. This is because, for a stationary planet, the outer ring develops (on a radial drift timescale for the dust) on the same timescale as the inner disc becomes devoid of dust (by the same process). 

\subsection{Role of the maximum grain size}
\label{subsec:max_dust}
\begin{figure*}
  \centering
  \includegraphics[width=0.8\textwidth]{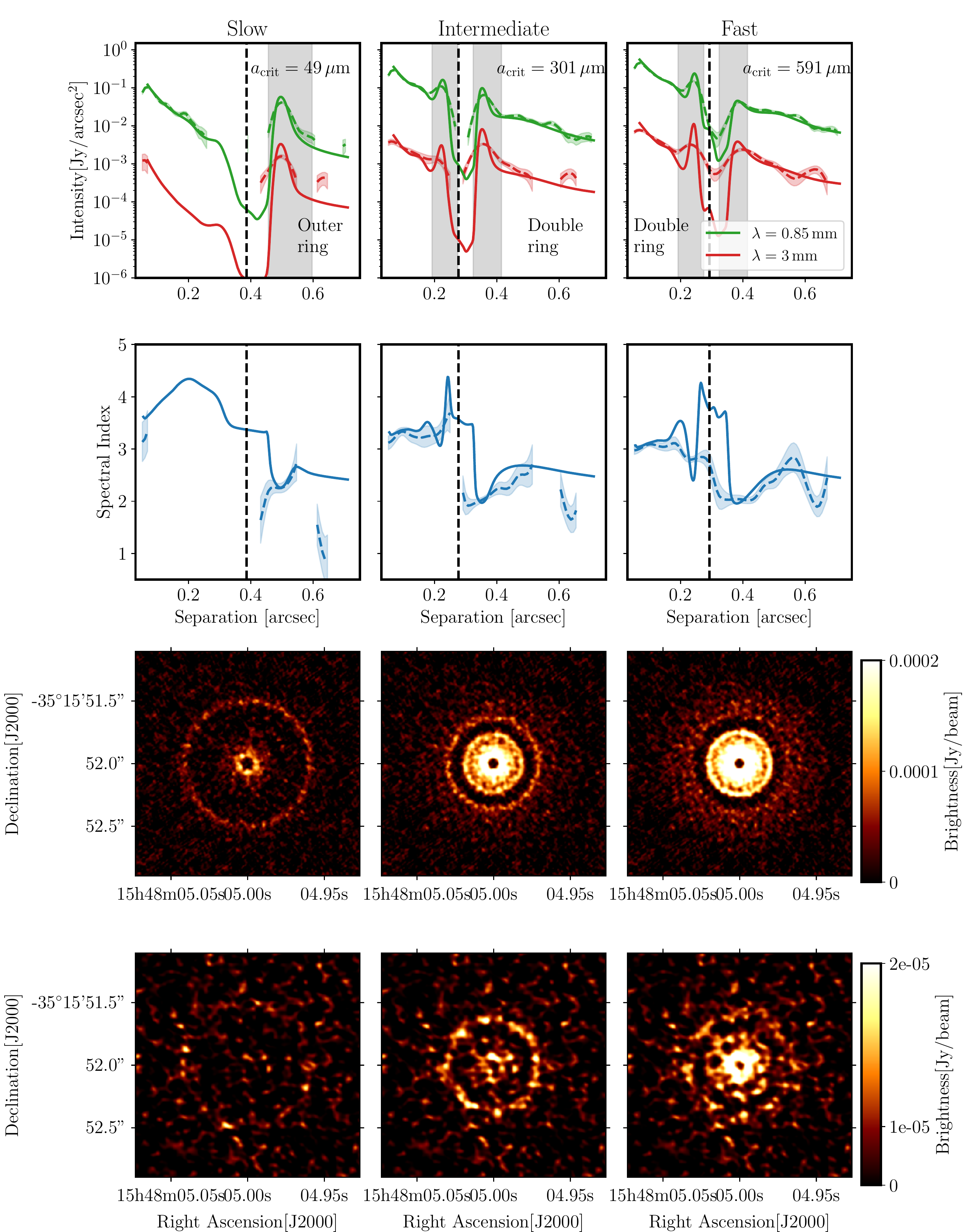}
  \caption{The same as \autoref{graph:multi_int} for $a_{\rm max} = 1\,{\rm cm}$. With the presence of larger grains, the intensity profile of the `Fast' simulation shows two maxima---i.e. this simulation produces structures similar to the `intermediate' simulation.}
  \label{graph:multi_int_1cm}
\end{figure*}

The main effect of changing the maximum grain size is to change the region of parameter space in which the planet is in the intermediate regime, characterised by two bright rings, one either side of the planet's orbit. This is demonstrated in \autoref{graph:multi_int_1cm}, which shows the intensity profiles for the same simulations as \autoref{graph:multi_int}, but with a maximum grain size of 1~cm instead of 1~mm. The morphology of the slow and intermediate cases are similar to those with $a_{\rm max} = 1\,{\rm mm}$, although the outer ring is noticeably brighter for $a_{\rm max} = 1\,{\rm cm}$. However, the morphology of the fast case has changed, showing a double ring. This is because the 1~cm grains are large enough that they migrate faster than the planet, becoming trapped in the pressure maximum outside the planet's orbit. Thus, for a larger $a_{\rm max}$ the fast planet simulation enters the intermediate regime.

The two sets of simulated images with $a_{\rm max} = 1\,{\rm mm}$ and $a_{\rm max} = 1\,{\rm cm}$ highlight the separate conditions required for the formation of the inner and outer rings. The outer ring forms as long as there are enough grains large enough that they migrate faster than the planet, even if these grains have sizes significantly larger than the wavelength. The transition disc-type morphology in the slow planet cases shows that the presence of grains that migrate slower than the planet is not sufficient to produce an inner ring, unless the remaining small grains provide a significant contribution to the opacity at $850\,\micron$. This means that the transition from the intermediate to slow regime will happen smoothly, with the intensity inside the planet's orbit decreasing as more grains inside the planet's orbit are able to escape onto the star. Despite this, the transition happens over a relatively narrow range in disc masses and migration speed because of the quadratic dependence of $a_{\rm crit}$ on the disc mass (\autoref{eqn:crit_a}).

\begin{figure*}
  \centering
  \includegraphics[width=0.8\textwidth]{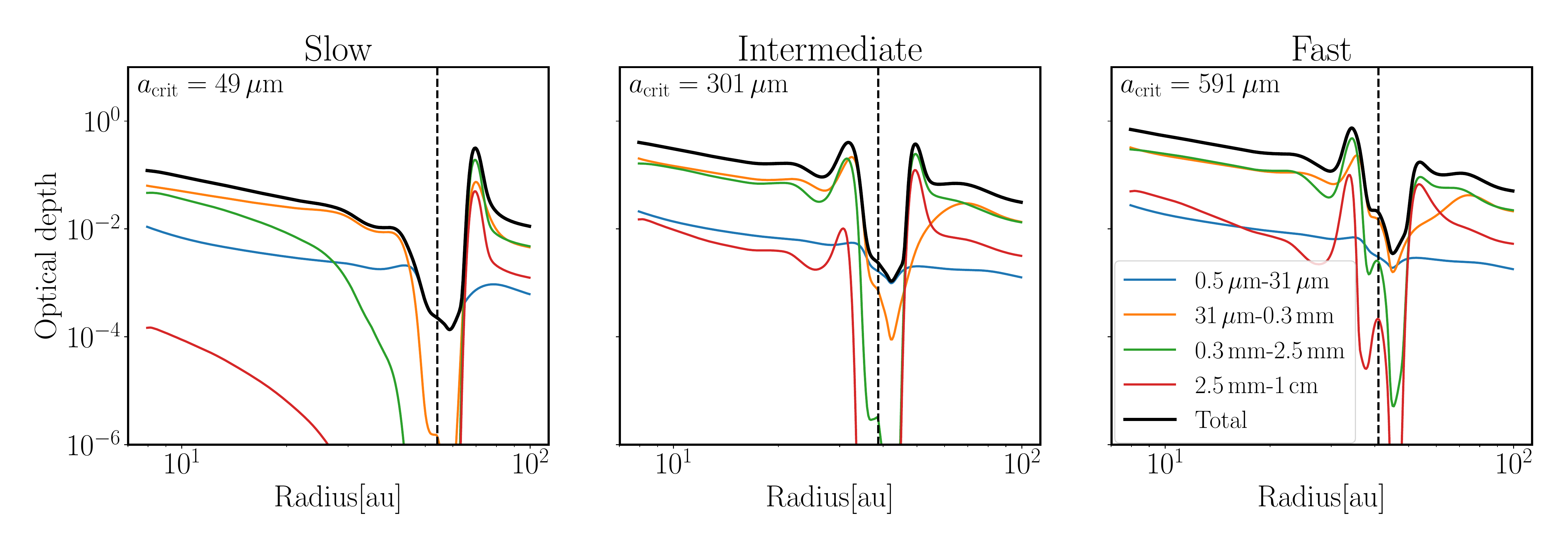}
  \caption{The optical depth calculated for four different dust bins at $850\,\micron$ for the models in \autoref{tab:mig}. The total optical depth at this wavelength is shown in black. The dashed line shows the location of the planet and the critical dust size is shown on each panel. This shows the contribution of the different dust grain sizes to the brightness profiles.}
  \label{graph:optical_depth}
\end{figure*}

To illuminate the above argument further, we break down the optical depth at $850\,\micron$ into the contributions from different grain sizes in \autoref{graph:optical_depth}. This shows that the outer ring is dominated by emission from grains of about $a_{\rm crit}$ and larger, while the emission from the inner ring is dominated by emission from grains with size comparable to $a_{\rm crit}$. Although the grains in the smallest size bin are predominantly found in the inner regions, their contribution to the total opacity is too small to be significant. Thus, the criterion $2 \upi a_{\rm crit} \lesssim \lambda$ can be used to approximately decide whether an inner ring should be present, distinguishing between the slow and intermediate cases. However, the criterion $a_{\rm max} \gtrsim a_{\rm crit}$ is most useful for determining whether an outer  ring forms because even the grains with $2 \upi a \gg \lambda$ contribute significantly to the opacity in the outer ring. We emphasise however that these criteria are `rules of thumb' and that whether a given feature, inside or outside the planet's orbit, would be detectable in practice would depend on details of the observational set-up (sensitivity and resolution).  Quantitative  interpretation of observations  thus requires bespoke simulations synthesised with matching observational parameters. 

\autoref{graph:optical_depth} can also be used to examine how the morphology would change in the case of smaller  $a_{\rm max}$. This is because only the normalization of the optical depth profile produced by a given bin of grain sizes is changed when the maximum grain size is changed (assuming all grains in the bin remain below the maximum grain size). The normalization of each bin can be computed simply because the optical depth scales with the normalization of the grain size distribution, which scales as $(1\,{\rm cm}/a_{\rm max})^{1/2}$ (for $n(a) \propto a^{-3.5}$). By looking at the optical depth profiles for grain sizes below 0.3~mm, we see that the intermediate simulation changes to be in the fast planet regime, while the slow planet simulation remains in the slow regime when $a_{\rm max} \lesssim 0.3\,{\rm mm}$. For the slow planet simulation case to transit to the fast regime, $a_{\rm max} \lesssim  a_{\rm crit}  = 40\micron$ would be required (\autoref{tab:mig}).

\subsection{Identifying migrating planets via spectral indices}
\label{sec:spec_index}
Now that we have shown that migrating planets produce different morphologies in the (sub)-millimetre continuum emission, we consider the issue of whether the signatures of planet migration can be unambiguously identified. We note that in order to \emph{quantify} the planet migration speed the radial drift velocity of the dust needs to be known, which in practice relies on an estimate of the gas surface density. However, in the absence of such an estimate, it is still possible to determine the velocity of the planet relative to the dust. 

The strongest signature of planet migration in the continuum emission would be to see the location of the ring change from inside the planet's orbit to outside it as the wavelength of the observations is increased. However, this is not seen in either \autoref{graph:multi_int} or \autoref{graph:multi_int_1cm}. A wider separation of wavelengths than $850\,\micron$ and 3~mm would be needed to see the maximum  change sides in a single disc. While going to shorter wavelengths (e.g. ALMA Band 9 or 10) would increase the leverage, obtaining high resolution data becomes more challenging due to the exceptional weather conditions required. Furthermore, the emission may become optically thick at short wavelengths. Wavelengths longer than 3~mm are currently not available with ALMA, and both high spatial resolution and high sensitivity would be required to see the effect, making such observations challenging. However, such long wavelength observations may be possible in the future with the ngVLA \citep{Ricci2018}.

However, even with ALMA observations at $850\,\micron$ and 3~mm it is possible to constrain the migration speed and distinguish between the signature of migrating planets and other mechanisms for producing gaps in some cases. 

First, in the case of a single ring outside the planet's orbit and a lower surface brightness inside the planet's orbit, the planet must be migrating slowly enough that $a_{\rm crit}$ is smaller than both $\lambda/2 \pi$ and $a_{\rm max}$ (see \autoref{subsec:max_dust}). This is expected to be the case with giant planets and transition discs. 

Second, we note the obvious case for a fast migrating planet, where there is only an inner ring. In this case we know that there cannot be any grains drifting faster than the planet migrates. Thus the planet's migration speed can be constrained by combining an estimate of the grain size outside the planet's orbit from the spectral index with a gas surface density measurement.

The intermediate case is more challenging, but offers the potential for the best constraint on the migration speed because each of the two rings limits the migration speed from different sides. In this case, we need to be able to distinguish the presence of two rings due to a migrating planet from two rings associated with another origin, such as two dust  traps formed by multiple non-migrating planets \citep[e.g.][]{Dipierro2015} or multiple gaps opened by a single planet in a low viscosity disc \citep[e.g.][]{Fedele2018}.

The spectral index can again be used as a diagnostic of a migrating planet in the intermediate case. The spectral indices are shown in \autoref{graph:multi_int} and  \autoref{graph:multi_int_1cm} (second row). The diagnostic feature is that the spectral index in the inner ring is higher than the outer ring, indicating that the grains are smaller in the inner ring. This is inevitable if the grain size in the inner ring is controlled by the competition between radial drift and planet migration, but is not expected for dust traps from multiple stationary planets. While we find that a higher spectral index on the inside is typically the case for the parameter space considered, we can envisage a situation where the spectral index appears to be low despite a fast migrating planet (i.e. a false negative). This could be the case if the rings become optically thick (although this is typically not the case for rings beyond $\sim 30\,{\rm au}$; \citealt{Andrews2018}). A second possibility is that the planet migrates so fast that it sweeps up large enough grains inside its orbit that the spectral index becomes low due to the presence of large grains. This would also require a grain size distribution with maximum size much greater than $\lambda$ to produce an outer ring. Empirically, we found that the optically thick condition is typically met first in our simulations. 

While it is possible for the spectral index in both rings to be low in the case of a migrating  planet, the scenario in which the inner ring has a higher index but there is no migrating planet is less likely (i.e. a false positive). This is for two reasons. 1) without any type of dust trapping the grains are expected to be larger closer to the star, i.e. the spectral index typically increases outwards. 2) If trapping is occurring in both rings, then large grains should be present in both of them. In fact, if growth is limited by fragmentation the grain size is expected to be larger in the inner trap \citep{Birnstiel2012,Bae2018b}, and thus have a lower spectral index in the inner trap, if there is a significant difference at all. Thus multiple rings due to multiple planets, or one planet opening multiple gaps in a low viscosity disc are unlikely to produce a structure with a lower spectral index in the outer ring.

Even in cases where the inner ring is not seen, such as due to the lower resolution in Band 3 (\autoref{graph:multi_int}), there is a clear difference in the surface brightness of the inner disc between the slow and intermediate cases. The slow case has a low surface brightness inside the gap, like a pre-transitional disc, while the intermediate case does not. Thus it may not be necessary to resolve the inner ring to determine that a planet is migrating. The presence of a ring outside the planet's orbit, combined with spectral index information showing that the grains are larger outside the gap than immediately inside of the gap would also confirm that the planet is migrating (regardless of whether there is a clear ring or just an enhanced surface brightness interior to the planet).

\subsection{Parameters of the disc}
\label{sec:parameters_of_disc}
\begin{figure}
  \centering
  \includegraphics[width=\columnwidth]{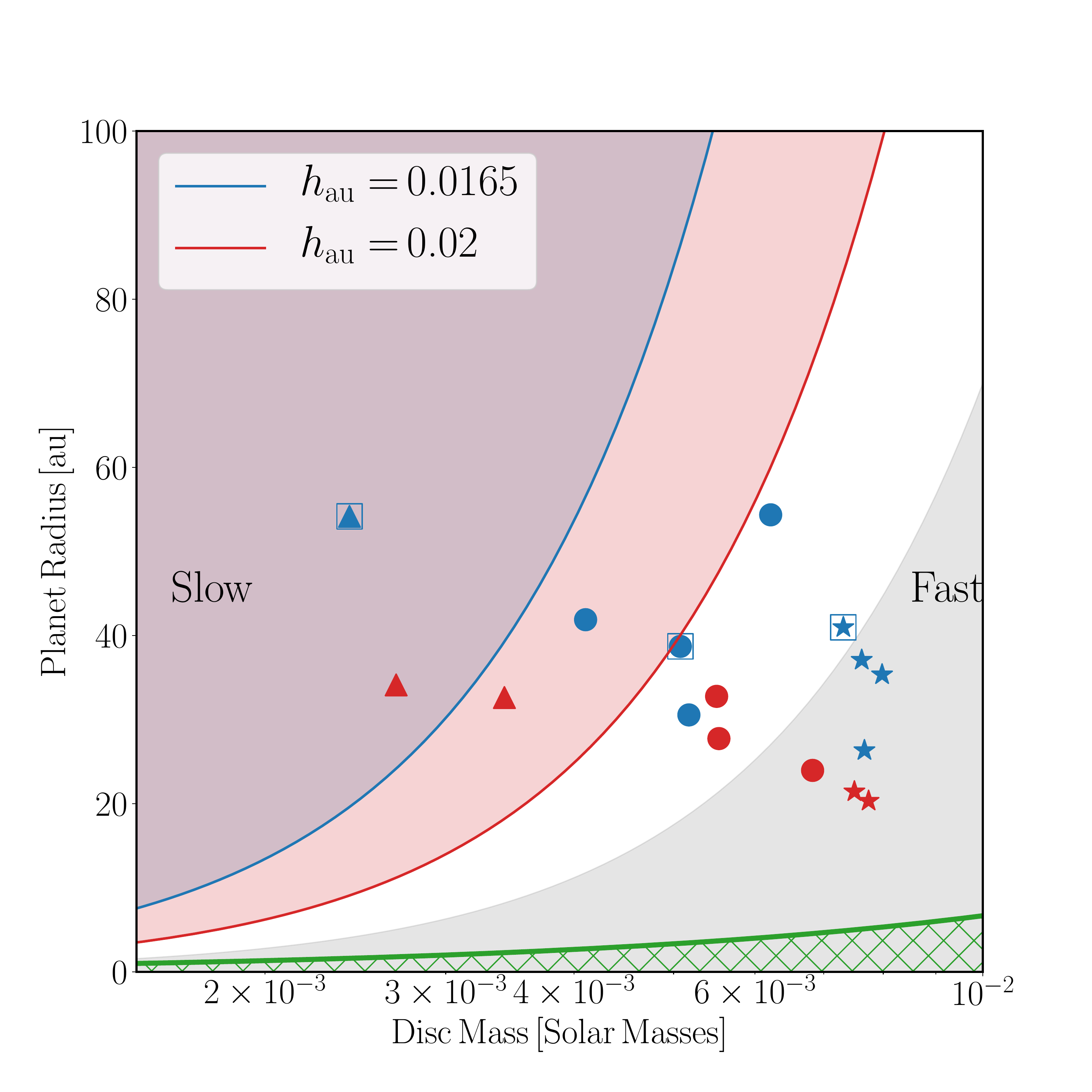}
  \caption{The parameter space in which we can observe migration of a $\rm 30\,M_{\oplus}$ planet. Colours denote models with different disc aspect ratios. Red and blue lines show where the migration speed of the planet is equal to the radial drift speed of dust grains with size $2 \upi a = 850\,\micron$. The shaded blue and red areas show where we get morphologies similar to the slow migrating planet (one ring outside planet's orbit). The shaded grey area shows where the models start becoming optically thick at $850\,\micron$. The green line and hatched area at the bottom show where $\rm St \leq \alpha_{visc}$ for $2 \upi a = 850\,\micron$ (i.e. where the dust follows the gas flow). Triangles show the simulations where a single ring forms outside of the planet's orbit (slow), circles show the simulations with two rings (intermediate) and asterisks show the simulations with a single ring inside of the planet's orbit (fast) for $a_{\rm max} = 1\,{\rm mm}$. The simulations discussed in detail in subsections \ref{first_order_intensity} to \ref{sec:spec_index} are shown with a box around them.}
  \label{graph:param}
\end{figure}

Having demonstrated that  different disc parameters give rise to noticeably different morphologies in the continuum emission associated with migrating planets, we now consider the parameter space in which the signatures of migration are present. For this purpose we consider a disc with $\Sigma \propto R^{-1}$ and $T \propto R^{-0.5}$, as in \autoref{sec:method}. In \autoref{graph:param} we explore this parameter space by showing the regions where a $30\,M_\oplus$ planet migrates faster than the dust that is the dominant contributor to the emission in Band 7, i.e. where $v_{\rm p} > v_{\rm d}$ for $2\upi a = 850\,\micron$. In this figure the disc mass refers to the gas mass inside 100~au (with a dust-to-gas ratio of 0.01). We also mark the approximate region where the disc becomes optically thick by the shaded grey area and where $\rm St \leq \alpha_{visc}$ for $2 \upi a = 850\,\micron$ by the green hatched area. On this figure we plot the results of a selection of different hydrodynamic results, with the derived morphology denoted by different symbols (divided into the categories above: a slow, intermediate and fast migrating planet).

From \autoref{graph:param}, the parameter range in which migration rate can be unambiguously measured appears to be quite small (i.e. the region between $2 \upi a_{\rm crit} = \lambda$ and the disc becoming optically thick). At low disc masses the planet migrates too slowly, resulting in a single ring appearing outside the planet's orbit (triangles). In this case it becomes difficult to distinguish a slowly migrating planet from a non-migrating planet. However, we note that it may still be possible to derive a constraint on the planet's migration speed from the fact that the inner disc has not been completely depleted of dust. Given enough time, for a non-migrating planet the surface brightness inside would decrease below the level found in \autoref{graph:multi_int}. Such an inference would however require detailed modelling of the dust evolution, because very small grains can filter through the gap \citep{Zhu2012, Pinilla2012,Bae2018b}.

At higher disc masses, a double ring structure in which the inner ring has a higher spectral index can be used to determine that a planet is migrating (circles). However, the spectral index diagnostic for migration can only be used until the disc becomes optically thick. This means that the  range over which it is possible to distinguish a single migrating planet from other mechanisms for producing a double ring structure is limited to a factor of a few in disc mass. However, for still larger disc masses where the speed of planet migration increases enough to cause the ring to switch to the interior of the planet (i.e. the asterisks in \autoref{graph:multi_int}), this  unambiguous morphological signature of migration does not require validation via spectral index measurements. Thus the regime in which it is possible to discern  planet migration extends into regions of  moderate optical depth (i.e. the grey shaded region in \autoref{graph:param}).

\section{discussion}
\label{sec:discussion}

In this paper we have investigated how rapidly migrating planets affect the morphology of the (sub)-millimetre emission and discussed the possible observable signatures of planet migration. \citet{Meru2018} identified that planet migration can significantly change the morphology of dust emission because it is possible for a planet to migrate faster than the dust. The simulated observations presented here show that the range of morphologies identified by \citet{Meru2018} are possible. For planets migrating slowly enough with respect to the dust, there is a prominent ring outside the planet's orbit, while for rapidly migrating planets a single ring appears inside the planet's orbit. For planets migrating at intermediate speeds, two rings are present. We find that the morphology can change from a single inner to a single outer ring over a relatively small range of migration speeds (a factor $\sim 4$ for the $a_{\rm max} = 1\,{\rm mm}$ case) because the size of the grains that migrate with the same speed as the planet, $a_{\rm crit}$, scales quadratically with the migration time-scale (due to the linear and quadratic dependencies on the disc mass in \autoref{eqn:typeImig} \& \ref{eqn:crit_a}, respectively). However, the change in morphology with wavelength is slower and  the morphology of the emission in ALMA Bands 3 and 7 are similar in each case. Thus it will be challenging to measure the migration speed by seeing the ring appear on different sides of the planet at different wavelengths.

\begin{figure}
  \centering
  \includegraphics[width=\columnwidth]{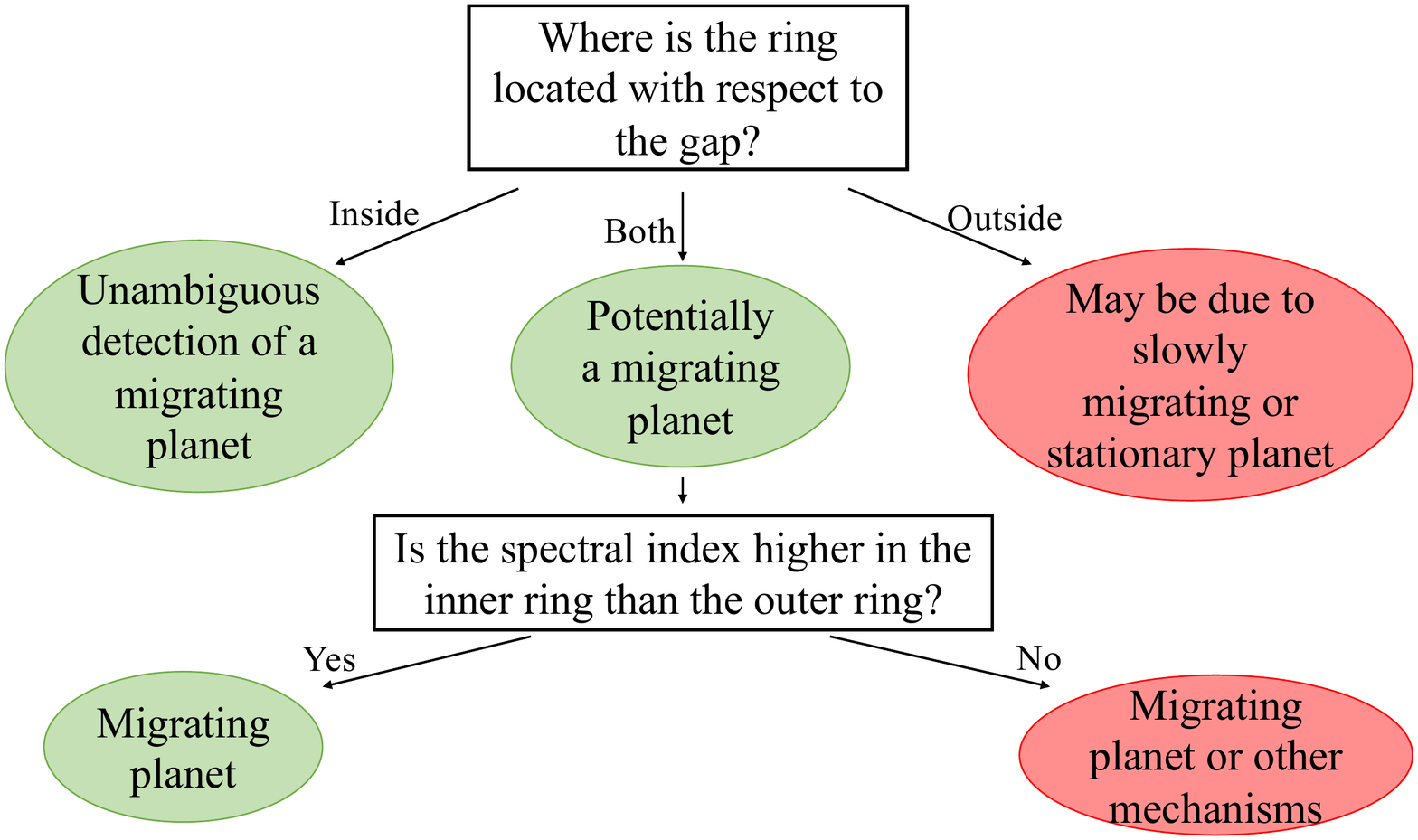}
  \caption{Summary of our findings. Green indicates when it is possible to detect a migrating planet while red indicates when it is not. Similar to the double ring case, a deep gap with an outer ring and a bright inner disc may also require a migrating planet.}
  \label{graph:flowchart}
\end{figure}

We suggest that the morphology and spectral index together can be used to constrain the planet's migration. \autoref{graph:flowchart} summarises our findings in a flowchart. The morphology of the emission at a single wavelength can place limits on migration when only a single ring is present, since we know the planet must be moving faster or slower than the dust (depending on whether the ring is inside or outside the gap, respectively). Similarly, when two rings are present we can infer that the planet is migrating faster than the smallest dust that contributes to the emission at the observed wavelength (typically 0.1~mm, \autoref{graph:optical_depth}), but slower than the largest dust present. The argument that the intermediate case constrains the migration speed between two limits relies on being able to distinguish two rings due to a migrating planet from other possible causes. In \autoref{sec:spec_index}, we argue that this is possible because the dust grains are larger in the outer ring and hence the spectral index there is lower. Conversely, for dust traps due to multiple planets or multiple gaps from a single planet in a low viscosity disc, the spectral index in the inner ring is generally expected to be lower.

\subsection{Do we already have evidence of migrating planets?}

Now that a number of protoplanetary discs have been observed at high resolution, it is interesting to consider whether the morphology of any of these discs provides evidence for planet migration. While the rings imaged in pre-transition discs are spectacular examples of dust trapping \citep[e.g.][]{Espaillat2014}, any planets responsible for the cavities in these discs are clearly in the slow or no migration regime.  It is therefore more promising to consider whether there are structures that correspond to the intermediate or fast regimes. The best candidates for this are discs with narrow rings and gaps; however, the two discs with spectral index maps at high resolution TW Hya and HL Tau \citep{Carrasco-Gonzalez2016,Huang2018a} do not have clear signatures of migrating planets. In HL Tau this may be because the emission in the ALMA bands is optically thick, while in TW Hya the gaps are shallow with no clear rings, suggesting any planets must have low masses \citep{Rosotti2016,Mentiplay2018} and thus allowing dust to flow through the gap.

The DSHARP survey \citep{Andrews2018} identified rings in $\sim 18$ discs, although not all of these are necessarily due to planets (\citealt{Zhang2016}, \citealt{Huang_snow2018}), and many of the gaps are shallow enough that dust could be flowing through the gap. Considering the systems with deep gaps, many of the discs show multiple deep gaps that are close together, such that the separate gaps and rings can not be considered independently. This makes it difficult to apply the arguments presented here without detailed modelling of both gaps simultaneously. Of the cases where there is a single deep gap, SR 4 and Elias 24, the argument for a migrating planet is strongest in the case of Elias 24. In the case of SR 4, the evidence for narrow rings either inside or outside of the planet's orbit is weaker, but any inference about the planet's migration must take into account the small size of the disc (i.e. since the gap is both close to the inner and outer edges of the disc, it is not obvious whether the planet could be in the fast or slow regime). 

Elias 24 shows a gap with a ring exterior to it. Although there is no clear second ring inside the gap, the disc's brightness is higher inside the planet's orbit than outside. This is difficult to reconcile with a non-migrating planet unless the planet is young, i.e. less than a radial drift time-scale old. However, such a young planet may be incompatible with the enhanced dust emission in the ring outside the gap. Thus we suggest that Elias 24 could host a planet migrating in the intermediate regime. We note that both \citet{Dipierro2018} and \citet{Zhang2018} have modelled Elias 24 with dusty hydrodynamical simulations. \citet{Dipierro2018} included migration and found good agreement with ALMA observations at 0.2\arcsec, while \citet{Zhang2018} used a non-migrating planet to model the DSHARP data. While the \citet{Zhang2018} results are a reasonable match to the data, the model overestimates the brightness of the ring outside the planet's orbit and underestimates the brightness inside the planet's orbit, thus we expect that including migration would likely lead to better agreement with the data. More robust conclusions as to whether the planet that is carving the Elias 24 disc is migrating or not can be drawn in two ways: first, with detailed hydrodynamical modelling that compares the predicted millimetre emission for migrating and non migrating planets. Second, through direct measurement of the spectral index made by comparing the current DSHARP observations (at 1.3mm) with high angular resolution observations at longer wavelengths (e.g., 3mm).

\section{Conclusions}
\label{sec:conclusion}

In this work we studied the possibility of observing the signature of a migrating planet, as predicted in \citet{Meru2018}. We conduct 2D dusty hydrodynamic simulations of planets migrating in protoplanetary discs and  produce synthetic ALMA continuum images at $\rm 850\,\micron$ and $\rm 3\,mm$ for a $\rm 30\,M_{\oplus}$ planet. 

We find that the brightness profiles of the dust continuum along with the spectral indices can constrain whether a planet is migrating, based upon three different morphologies for the brightness profiles. The most radical differences between the brightness profiles occurs between a slowly and a fast migrating planet. For a planet migrating slower than the dust, a dust ring forms outside of the planet's orbit, whereas if the planet migrates faster than the dust the ring appears inside of the planet's orbit. We argue that these morphologies can be used to constrain the planet migration speed, if the radial drift velocity of the dust can be estimated, for example via the spectral index of the millimetre emission.

The third morphology occurs for planets migrating with an intermediate speed. These planets create two rings with largest dust grains being trapped in the outer ring and smallest dust grains being swept into an interior ring by the migrating planet. In this case, we can use the spectral indices to distinguish between a migrating planet and other means of creating multiple rings (i.e. low viscosity discs or multiple planets). For a migrating planet we expect to see a lower spectral index for the outer ring created by the larger dust grains that have been trapped there.

Finally, we suggest that of the discs observed at high angular resolution, Elias 24 is the most promising case for a single planet migrating at a speed comparable to the dust, due to the presence of a ring outside of the gap and an enhancement of dust interior to the planet.                 

\section*{Acknowledgements}

This work has been supported by the DISCSIM project, grant agreement 341137 funded by the European Research Council under ERC-2013-ADG. This work was performed using the DiRAC Data Intensive service at Leicester, operated by the University of Leicester IT Services, which forms part of the STFC DiRAC HPC Facility (www.dirac.ac.uk). The equipment was funded by BEIS capital funding via STFC capital grants ST/K000373/1 and ST/R002363/1 and STFC DiRAC Operations grant ST/R001014/1. DiRAC is part of the National e-Infrastructure. G.R. acknowledges support from the Netherlands Organisation for Scientific Research (NWO, program number 016.Veni.192.233). F.M. acknowledges support from the Royal Society Dorothy Hodgkin Fellowship. We also thank the referee for a constructive report.




\bibliographystyle{mnras}
\bibliography{bib_file} 


\bsp	
\label{lastpage}
\end{document}